\documentclass[final,5p,times,sort&compress,twocolumn]{article}

\usepackage[separate-uncertainty=true,list-units=repeat,multi-part-units=brackets]{siunitx}

\usepackage{amssymb}
\usepackage{amsmath}
\usepackage{subcaption}
\usepackage{graphicx}
\usepackage{caption}
\usepackage{media9} 
\usepackage{float}
\usepackage{booktabs}
\usepackage{multirow}
\usepackage{booktabs}
\usepackage{multirow}
\usepackage{threeparttable} 
\usepackage{microtype}
\usepackage{multirow}
\usepackage{booktabs}  
\usepackage{svg}
\usepackage{tikz}
\begin{document}




\title{Resolving Discrepancies in Wood Micromechanics: Strain-Mapped Compression of Tracheid Wall Micropillars}

\author{
  Júlio O. Amando de Barros\textsuperscript{1,2},
  Jakob Schwiedrzik\textsuperscript{2,3},
  Falk K. Wittel\textsuperscript{1} \\[1ex]
  \small \textsuperscript{1}Institute for Building Materials, ETH Zurich,\\
  \small Laura-Hezner-Weg 7, 8093 Zurich, Switzerland\\
  \small \textsuperscript{2}Laboratory for Mechanics of Materials and Nanostructures, Empa,\\
  \small Feuerwerkerstrasse 39, 3602 Thun, Switzerland\\
  \small \textsuperscript{3}Laboratory for High Performance Ceramics, Empa,\\
  \small Überlandstrasse 129, 8600 Dübendorf, Switzerland
}

\twocolumn[
\begin{@twocolumnfalse}
\maketitle

\begin{abstract}
\vspace{-1em}
Wood’s increasing role as a structural resource in sustainable materials selection demands accurate characterization of its mechanical behavior. Its performance arises from a hierarchical structure, where the dominant load-bearing component is the S2 layer of tracheid cell walls—a thick, fiber-reinforced composite of cellulose microfibrils embedded in hemicelluloses and lignin. Due to the small dimensions and anisotropic nature of the S2 layer, mechanical testing presents significant challenges, particularly in producing homogeneous stress and strain fields. In this study, we apply micropillar compression (MPC) combined with digital image correlation (DIC) to Norway spruce tracheids, enabling direct and model-free strain measurements at the cell wall scale. Micropillars were oriented at different microfibril angles (MFAs), confirming the expected dependence of stiffness and yield stress on ultrastructural alignment, with higher stiffness and yield stress at low MFAs. For these under compression fibril-aligned kink bands occurred, while shear related failure was observed at higher angles. A parameter study on the acceleration voltage of the Scanning Electron Microscope revealed that electron beam exposure significantly degrades pillar integrity, which could explain data scatter and mechanical underestimation in earlier MPC studies. By controlling imaging protocols and using DIC-based strain measurements, we report the highest direct measurements of wood cell wall stiffness to date—up to \SI{42}{\giga\pascal} for MFA=\SI{0}{\degree}—closer matching micromechanical model predictions compared to previous results. Findings are compared with Finite Element Method-based displacement corrections to establish a robust protocol for probing soft, anisotropic biological composites' mechanical behavior while clarifying longstanding inconsistencies in reported results of wood MPC measurements.
\vspace{1em}
\end{abstract}
\end{@twocolumnfalse}
]

\section{Introduction}
\label{sec1}

Wood is a common but increasingly important renewable, carbon-storing resource with exceptional specific stiffness and strength, ideal as a construction material in the transition to a greener future~\cite{jiang2018wood,farid2022transforming}. Softwood species, such as Norway spruce (\textit{Picea abies}), are currently the dominant construction wood and are the focus of this study. The macroscopic performance of wood as a cellular, porous material stems from its hierarchical structure~\cite{gibson2012hierarchical, amando2024unifying}. At each scale, evolution selected natural composites: on the macroscopic scale, softwood is formed by growth rings, alternating from early, to transition, and latewood~\cite{bertaud2004chemical,ferrara2025tensile}; each of these tissues are formed on the mesoscale by parallel arrangements of tracheids interconnected by the middle lamella, together with other structures such as ray tissue and resin channels; on the microscale tracheids are elongated, fiber-like cells, with their cell wall consisting of a primary  (P) and a secondary (S) wall, arranged concentrically around the cell lumen; the S wall is, in principle, a laminate formed by three secondary layers (S1, S2, S3). The secondary layers themselves can be regarded as a unidirectional fibrillous composite with cellulose fibril aggregates (CFA) with a characteristic fibril angle with respect to the tracheid axis, known as the microfibril angle (MFA)~\cite{barnett2004cellulose}. The CFAs are, in fact, a composite themselves, formed of cellulose microfibrils and hemicelluloses, embedded in hemicelluloses and lignin ~\cite{boyd1982anatomical}. The low MFA and high CFA content of the S2 layer, together with its great thickness, are the reasons why it is regarded as most relevant for all mechanical analyses and also stands in the focus of this study~\cite{sjostrom1999chemical,bergander2002cell,fengel2011wood,salmen2004micromechanical,salmen2009cell,engelund2013critical,stevanic2020molecular}.

The synthesized biopolymers that form the cell wall are polysaccharides like crystalline and amorphous celluloses, hemicelluloses, and aromatic polymers like different kinds of lignin. It is important to note that CFAs comprise aligned chains of strong and stiff cellulose—linear polymers of D-glucose units connected by $\beta(1\rightarrow4)$ glycosidic bonds. The high density of hydroxyl groups leads to extensive intra- and intermolecular hydrogen bonding, resulting in a semi-crystalline structure. The combination of strong covalent and weak hydrogen bonding interactions results in a high stiffness, making CFAs the primary stiffening component of the cell wall~\cite{fahlen2002lamellar,donaldson2019wood}. In contrast, hemicelluloses are a heterogeneous group of branched polysaccharides, and form together with lignin, that are complex and irregular aromatic polymers derived from phenylpropanoid, the amorphous matrix~\cite{boerjan2003lignin,scheller2010hemicelluloses}. Any ion or electron beam used for preparation or observation inevitably interacts on the molecular level with the possibility to excite, alter, or degrade the biopolymers of wood to a great extent~\cite{jakob2017femtosecond}. Nevertheless, many experimental techniques on the microscale rely on electron microscopy. 

Mechanical experiments on the length scale of the cell wall pose several significant challenges. In spruce, the cell wall thickness of latewood tracheids is in the order of \SIrange{5}{6}{\micro\meter}, while earlywood walls are only \SIrange{1}{2}{\micro\meter} thick. Note that about \SI{90}{\percent} of the thickness corresponds to the S2 layer~\cite{fahlen2005cell}. The most popular mechanical testing techniques at this scale are Nanoindentation~\cite{wimmer1997longitudinal,konnerth2009actual,jager2011relation,bertinetti2015characterizing} and Atomic Force Microscopy~\cite{clair2003imaging,zimmermann2006ultrastructural,casdorff2017nano}. Both were used to reveal the cell wall structure by probing the mechanical response of individual layers or mapping the distribution of CFAs within the matrix with a sub-micrometer resolution. However, these techniques inherently generate non-trivial stress fields during indentation of the orthotropic material, complicating the interpretation of the mechanical response. As a result, the material parameters must be obtained inversely by models, which are often oversimplified and exclude key information, such as anisotropy or edge effects. This results in stiffness values that tend to be significantly underestimated compared to state-of-the-art micro-mechanical cell wall models~\cite{salmen2004micromechanical,casdorff2017nano,gindl2004significance,mora2019mechanical}. Experimental approaches that impose cleaner stress states and an accurate alignment between materials and indentation orientations are highly desirable for accurate mechanical characterization.

The realization of a clean, uniaxial stress state in the test volume is the most critical and challenging task to improve micro-mechanical testing. In this respect, tests on microsamples, which can be produced by focused ion beam (FIB) milling, are the most promising~\cite{reyntjens2001review}. Free-standing pillars for compression with flat-punch microindentors have proven to be the most reliable approach for materials characterization at the micrometer scale~\cite{hemker2007microscale, jayaram2022small}. From the recorded force and displacement, material parameters can be calculated, at least in principle. Despite its promise, those micropillar compression (MPC) tests face several challenges, such as ensuring accurate pillar geometry, minimizing beam-induced material modification, mitigating sink-in effects into the substrate, and identifying the onset of mechanical contact~\cite{zhang2006design,konstantinidis2016correct,cornec2022numerical} as well as barrelling effects due to high friction with the indentor. To overcome these limitations, modeling-based corrections are often employed~\cite{zhang2006design}. Since the 2010s, MPC testing has been applied to wood for obtaining material parameters such as Young’s modulus and yield stress~\cite{zhang2010characterizing,schwiedrzik2016identification,klimek2020micromechanical} and to characterize failure patterns on the cell wall scale~\cite{adusumalli2010deformation,raghavan2012deformation}. Unfortunately, the reported wood MPC data exhibit large scatter and still underestimate stiffness values to a considerable extent, compared to expected values from micromechanical models. This calls for a comprehensive investigation of MPC tests on wood, encompassing the entire process from raw material preparation to final data interpretation. Possible error sources need to be identified, and the role of the test volume orientation with respect to the MFA needs to be clarified.  

In this work, MPC tests on Norway spruce are enhanced by applying digital image correlation (DIC) on microdots that were deposited on the micropillar surfaces. The comparison with strains calculated from the indentor displacement allows for a reinterpretation of the MPC test for wood. With the improved MPC protocol, the orthotropic nature of the S2 layer is assessed by varying micropillar orientations. Via in-situ tests inside the scanning electron microscope (SEM), the evolution of orientation-dependent strain localizations and sink-in of the elastic foundations are observed. By varying the electron beam scanning protocols, the role of beam-wood interactions as a decisive parameter for reproducibility is identified. The paper is organized as follows: First, the micropillar preparation, testing protocols, and data analysis procedures are described, followed by a study of the effects on micropillar quality and the role of electron beam damage. The new measurements are compared to previous ones, demonstrating that the significant scatter reported in the literature can be attributed to undetected beam damage. Finally, the potential and current limitations of MPC tests for soft organic biocomposites are discussed.

\section{Materials and Methods}\label{sec2}
\begin{figure*}[t]
    \centering
    \includegraphics[width=1\linewidth]{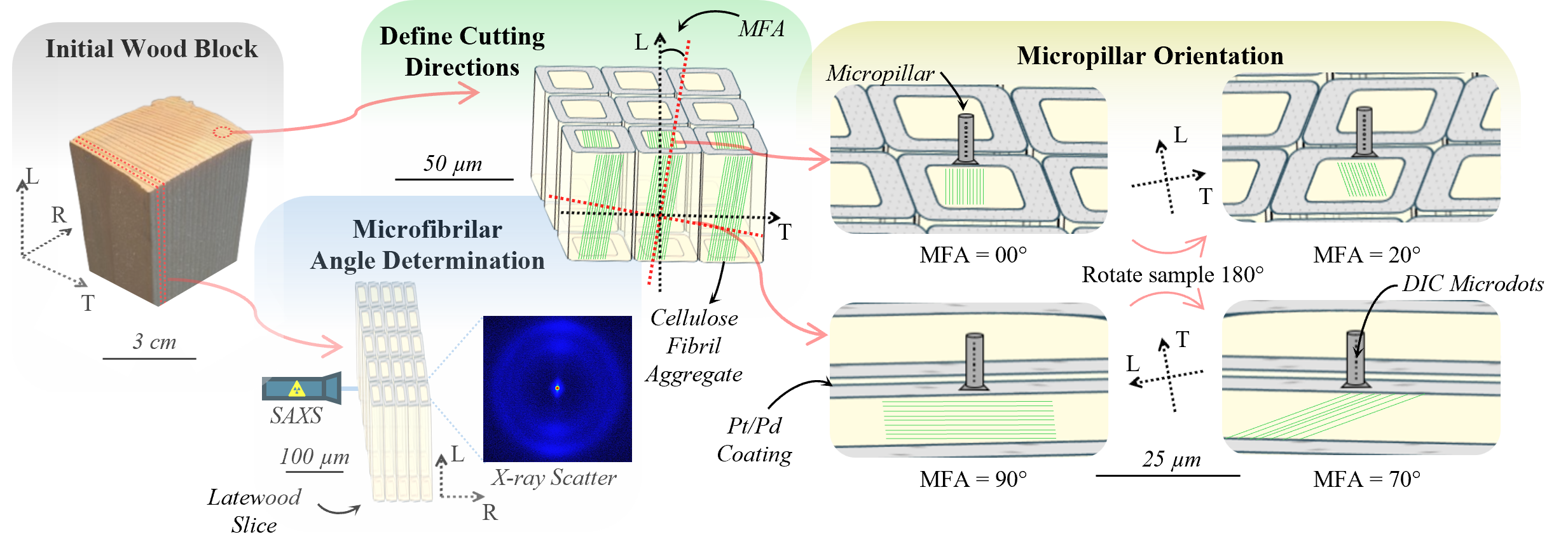}
    \caption{Sample preparation scheme: From a wood block, flat and cylinder-shaped samples are extracted for MFA determination and micropillar preparation, respectively. Cylinders have two varying orientations: One for MFA=\SI{0}{\degree} and \SI{20}{\degree}, and a perpendicular one for \SI{90}{\degree} and \SI{70}{\degree}. After FIB milling, S2 micropillars remain and get marked by microdots via deposition.} \label{fig:sampleprep}
\end{figure*}

\begin{figure}[ht]
\begin{center}
\begin{minipage}[b]{0.48\textwidth}
    \centering
    \includegraphics[width=1\textwidth]{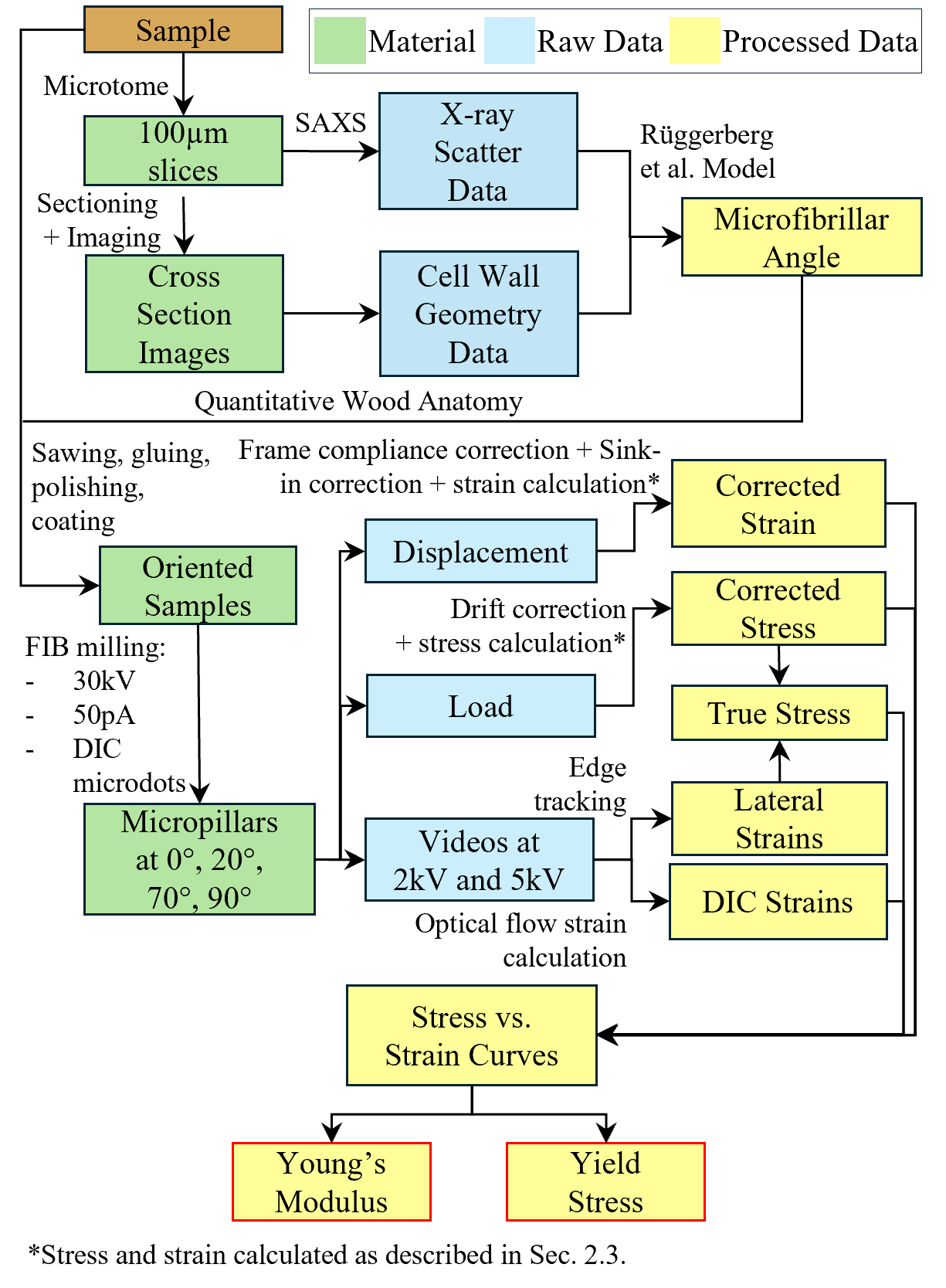}
    \end{minipage}
    \caption{Flow diagram of procedures from sample preparation to data analysis.}
    \label{fig:flowdia}
\end{center}
\end{figure}

Micromechanical testing on micropillars comes along with a chain of interdependent challenges, including processing, preparation, actual measurements, data analysis, and interpretation, where errors are difficult to detect, due to the extremely small dimensions of the samples. This materials and methods chapter is organized into sections for sample preparation (Sec.\ref{sec:Pillarprep}), the actual testing (Sec.~\ref{sec:compprot}), and the data analysis (Sec.~\ref{sec:dataanal}). While the sample preparation is schematically shown in Fig.~\ref{fig:sampleprep}, the entire experimental protocol is summarized in the flow diagram of Fig.\ref{fig:flowdia}. 
\subsection{Micropillar Preparation}\label{sec:Pillarprep}
All samples were prepared from a Norway spruce (\textit{Picea abies}) tree grown in Alpthal, Switzerland, at an altitude of \SI{1093}{\meter}. A log section was collected between \numrange{1}{2}\SI{}{\meter} above ground and sectioned in the green state to prevent cracking during drying at a \SI{65}{\percent}RH/\SI{20}{\celsius} climate. After drying, a longitudinal beam with dimensions $3 \times 3 \times 50$\SI{}{\cubic \cm} (radial (R) $\times$ tangential (T) $\times$ longitudinal (L)) was cut approximately \SI{ 15}{cm} from the pith. Specimens were extracted from this beam for the determination of the microfibril angle (MFA) and for all micropillar compression tests (see Fig.~\ref{fig:sampleprep}).

Knowing the MFA is crucial for orienting the local S2 cell wall fibril orientation with the indentation direction. For this purpose, the MFA was determined with the enhanced X-ray scattering method developed by Rüggeberg at al.~\cite{ruggeberg2013enhanced}. Three transverse slices ($\approx$\SI{100}{\micro\meter} thick) were used to obtain the scattering patterns (see Fig.~\ref{fig:sampleprep}). The (200) scattering of crystalline cellulose was integrated over the azimuthal angle to determine the predominant orientation of the cellulose fibrils relative to the incident beam. The test sample was subsequently sectioned, and the anatomical orientations and extensions of the tracheid walls were measured. By combining anatomical, crystalline, and X-ray data, the most probable MFA was calculated to be \SI{10}{\degree} \cite{ruggeberg2013enhanced}, which agrees with values previously reported~\cite{brandstrom2001micro}.

Based on the determined MFA, the wood beam was further cut at angles of \numlist{10;100}\SI{}{\degree} relative to the tracheid axis (L-direction). This corresponds to fibril orientations perpendicular and parallel to the surface, respectively (see Fig.~\ref{fig:sampleprep}). Note that the cutting plane is perpendicular to the LT-plane, to ensure that the pillars are milled from tangential walls with reduced pit density. Due to the helicoidal arrangement of cellulose fibrils, only one tangential wall per tracheid aligns precisely with the target orientation while the opposite one is misaligned by $2 \times \mathrm{MFA}$. Consequently, the two selected cutting angles resulted in cell walls oriented at MFAs of \numlist{0;20}\SI{}{\degree}, as well as \numlist{70;90}\SI{}{\degree} (see Fig.~\ref{fig:sampleprep}).

Cylinders with a diameter of about \SI{6}{mm} and \SI{2}{mm} in height were extracted from the oriented samples and glued on standard \SI{12}{mm} SEM stubs with a fast-setting highly viscous 2K-epoxy-based glue (UHU plus). After curing of the adhesive, the wood surfaces were polished with a rotary microtome at a step size of \SI{20}{\micro\meter} (Leica RM2255). The polished surfaces were then coated with a conductive \SI{10}{\nano\meter} Platinum/Paladium layer to enhance imaging quality under the SEM by reducing surface charging. 

Micropillars were milled from the cell walls by using a gallium Focused Ion Beam (FIB) at an acceleration voltage of \SI{30}{\kilo\volt}. A single-step milling protocol was used to minimize ion-induced damage with a low current of \SI{50}{\pico\ampere}. The milling time was approximately \SI{15}{\minute} per pillar. The resulting pillars were, on average, \SI{5}{\micro\meter} in height, with a length/diameter aspect ratio of 2. Finally, a row of platinum microdots was deposited on the pillar surface using Focused Electron Beam Deposition (FEBID) for the strain field measurements via DIC. The final deposition protocol employed parallel scan mode, a low dwell time of \SI{0.5}{\micro\second}, and a deposition duration of approximately \SI{0.5}{\second} to avoid beam damage. In total, 15 pillars were prepared per orientation. 

\begin{figure}[ht]
    \centering
    \includegraphics[width=1\columnwidth]{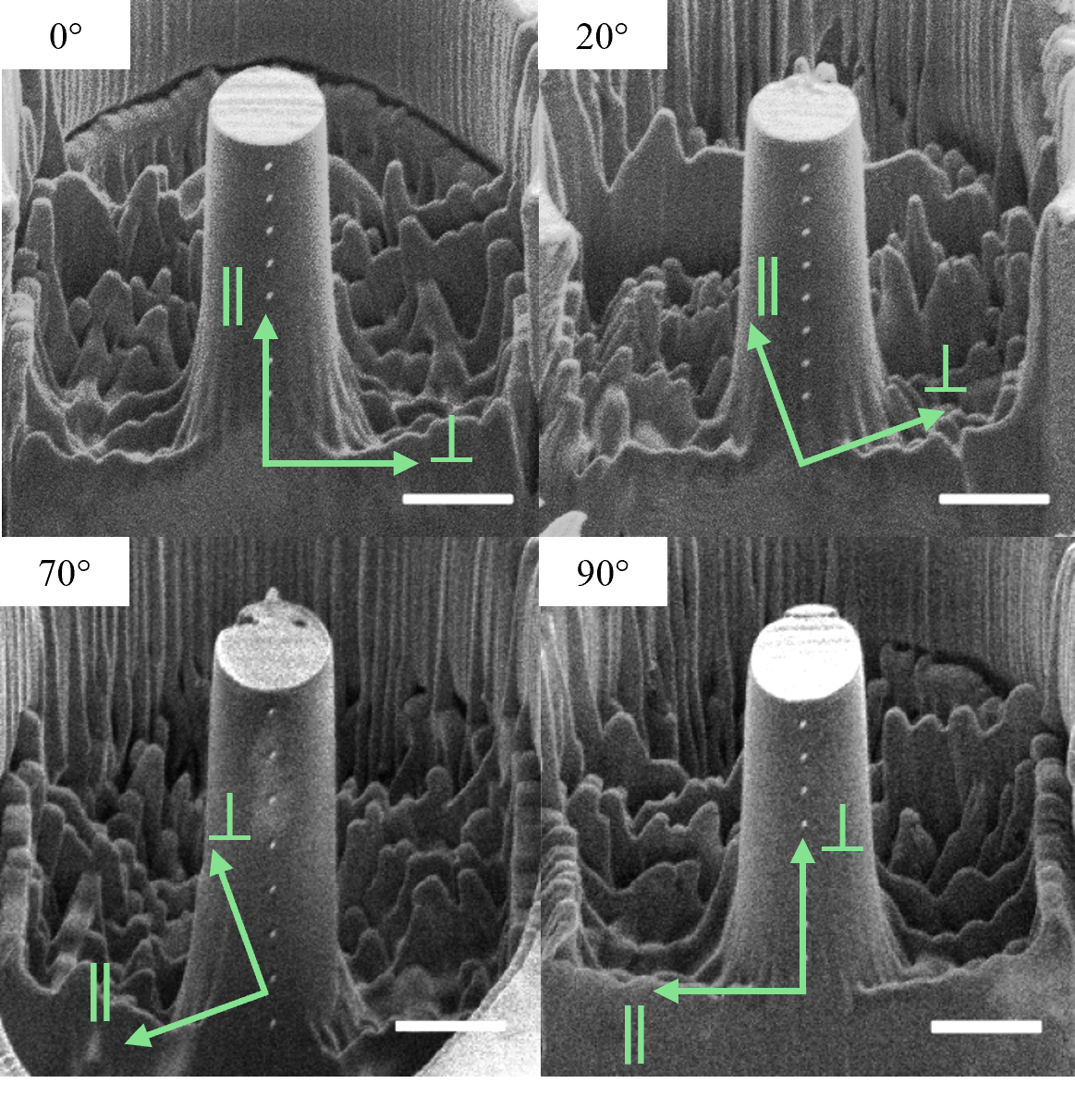}
    \caption{SEM images of produced pillars on different microstructural orientations before testing. Scale bars represent $2 \mu m$.}
    \label{fig:pillars_pre}
\end{figure}

\subsection{Compression protocol}\label{sec:compprot}
\begin{figure}[ht]
    \centering
    \includegraphics[width=1\columnwidth]{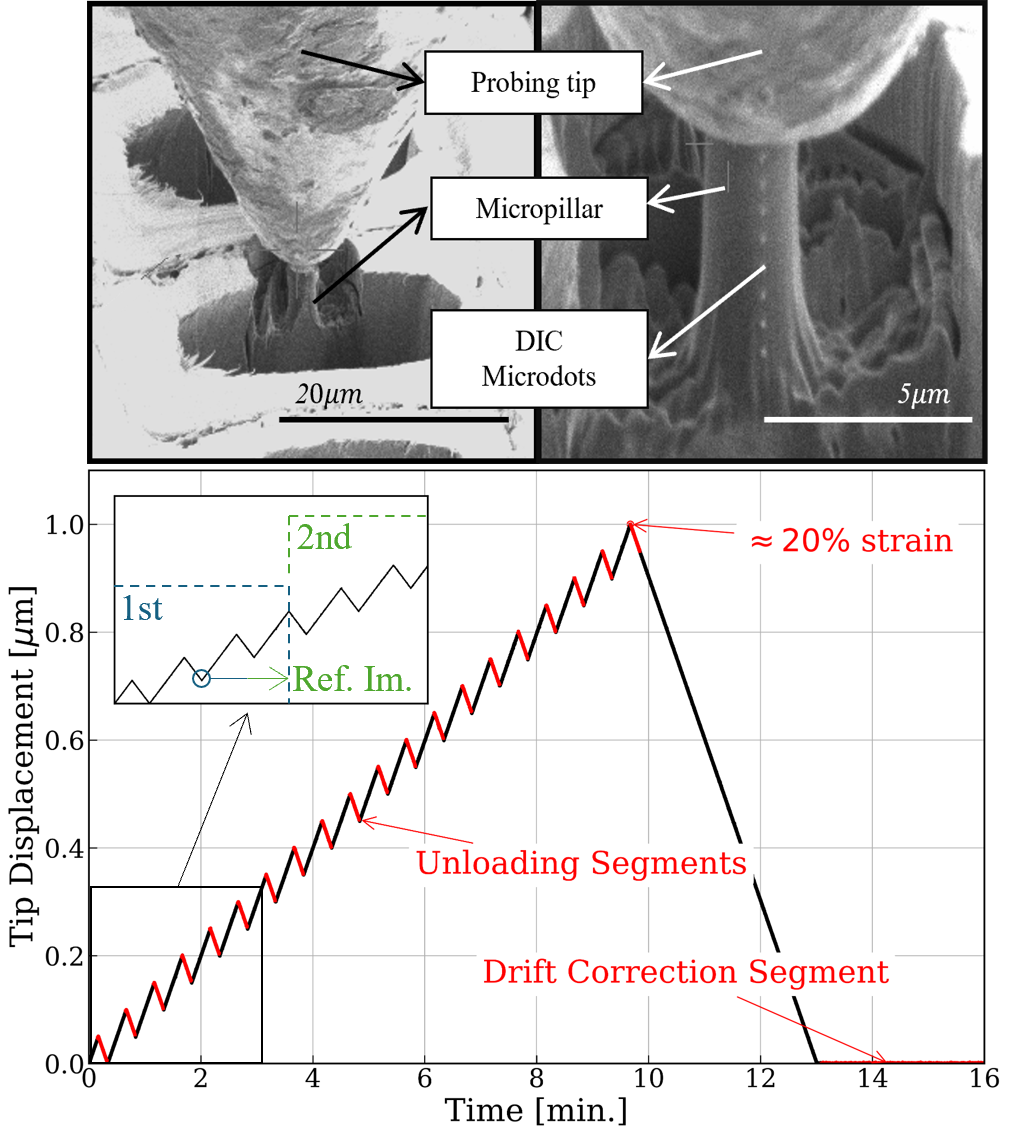}
    \caption{Top: Test setup with tip and extracted micropillar with microdots on the cell wall at low scanning magnifications for test setup (l) and high one for testing purposes (r). Bottom: Loading history for the tip displacement control with inset showing the subsets for DIC strain calculation.}
    \label{fig:compression_prot}
\end{figure}
All micropillar compression tests were performed in displacement control inside a SEM under high vacuum conditions (Tescan Mira 3) with a \SI{5.5}{\micro\meter} flat punch microindentor (Alemnis) and a \SI{0.5}{\newton} load cell (Honeywell). Imaging was conducted under three scanning conditions to assess the effect of beam damage: (i) single image acquisition every two minutes (referenced as "\textbf{No-beam}"), (ii) continuous scanning at an acceleration voltage of \textbf{\SI{2}{\kilo\volt}}, and (iii) continuous scanning at \textbf{\SI{5}{\kilo\volt}}. A beam current of \SI{50}{\pico\ampere} was used in all cases. During all preparation steps, low-magnification scans ($\leq$ \SI{2000}{\times}) are used (see Fig.~\ref{fig:compression_prot} top left) to minimize beam damage, while images suitable for DIC require a higher magnification (\SI{8000}{\times}) (see Fig.~\ref{fig:compression_prot} top right), which leads to higher material degradation.

To minimize thermal drift, temperature changes, and outgassing of residual moisture, the mounted set-up was kept inside the vacuum chamber overnight before testing. A maximum compressive displacement of \SI{1}{\micro\meter} was incrementally applied at a strain rate of \SI{1e-3}{\per\second}, corresponding to a total strain of approximately \SI{20}{\percent} (see loading history in Fig.~\ref{fig:compression_prot}, note that compressive strains are defined as positive throughout the paper for visualization purposes). Multiple partial unloading segments were included, each with a strain decrement of approximately \SI{1}{\percent}, to capture the material's elastic response. Each unloading segment yielded ten images suitable for DIC strain analysis at a frame rate of \SI{1}{\hertz}. The typical test lasted approximately \SI{16}{\minute}, including waiting times for a drift correction.
\subsection{Data analysis}\label{sec:dataanal}
\begin{figure}[ht]
    \centering
    \includegraphics[width=1\columnwidth]{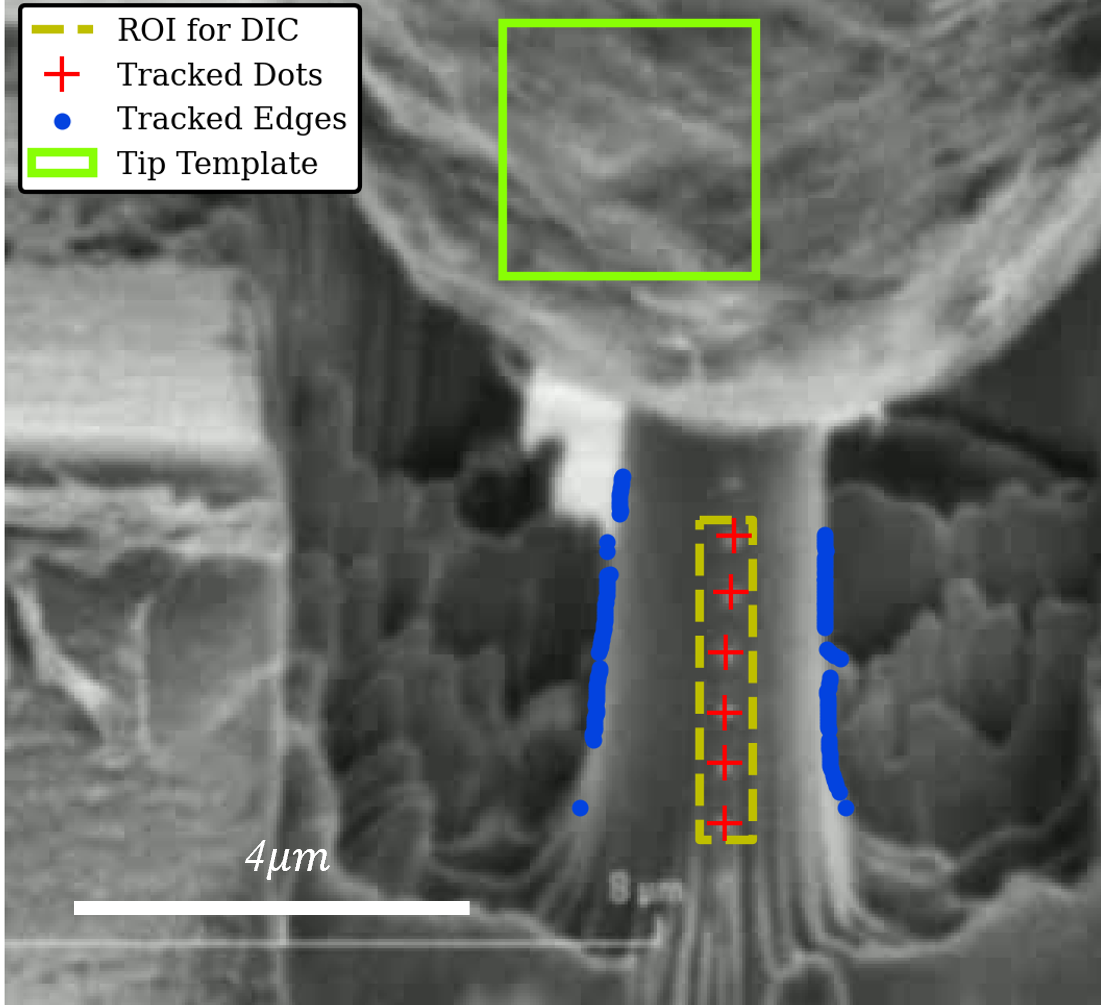}
    \caption{Example image with tracked features. The yellow dashed rectangle indicates the region of interest (ROI) used for digital image correlation (DIC) strain computation. Red crosses mark tracked reference points used to update the ROI in each image. Blue dots highlight the detected edges of the pillar. The green square shows the tip template used to synchronize the DIC data with the load data.}
    \label{fig:dic_strain}
\end{figure}

Since this work relies on comparative image-based strain measurements, special attention was given to robust image analysis routines for DIC. For each image sequence, the deposited microdots, the pillar edges, and the indentor were tracked (see Fig.~\ref{fig:dic_strain}). Additionally, the noise was reduced with a rolling average filter with a Gaussian kernel that comprises three consecutive images. Due to large displacements and deformations, as well as shading effects, the use of a single initial reference image for DIC strain calculations is not advisable. Hence, an incremental strain calculation is applied on consecutive image subsets. Since images close to the reference image of the subset only have small relative strains, a low signal-to-noise ratio would follow. To avoid such noisy data from entering the total strain calculation, the reference image for the image subset is taken at half of the previous one. When needed for robustness, the region of interest (ROI) was adapted to the actual situation of the image subsets using the \textbf{tracked microdots} (dashed rectangle in Fig.~\ref{fig:dic_strain}). Displacements within the ROI were computed with the Farnebäck optical flow algorithm, as implemented in OpenCV \cite{farneback2003two,bradski2000opencv}. From the ROIs, the mean strain increments were calculated and cumulated to obtain the total time-strain curve. For the \textbf{micropillar edges recognition}, independent ROIs were defined around the left and right edges. In each ROI, horizontal intensity profiles were computed at every row by averaging a vertical window of five consecutive rows centered at that position (e.g., from $i-2$ to $i+2$). The horizontal gradient of this 1D profile was then computed using the first derivative to highlight changes in intensity. To segment the pillar from the background, Otsu’s thresholding method~\cite{otsu1975threshold} was applied. The resulting detected edges are shown in Fig.~\ref{fig:dic_strain}. To reduce noise in the lateral strain calculation, only edge positions within a vertical window centered at the pillar’s mid-height were considered. Additionally, points deviating significantly from the mean horizontal edge position were excluded to filter out incorrectly detected edges, see Fig.~\ref{fig:dic_strain}. From the images, also the indentor displacement can be obtained by \textbf{indentor tracking}, using a rectangular template for template matching as implemented in OpenCV with normalized cross-correlation coefficient~\cite{bradski2000opencv}. Its displacement was used to synchronize the DIC data with the measured load-displacement data.

\begin{figure}[ht]
    \centering
    \includegraphics[width=1\columnwidth]{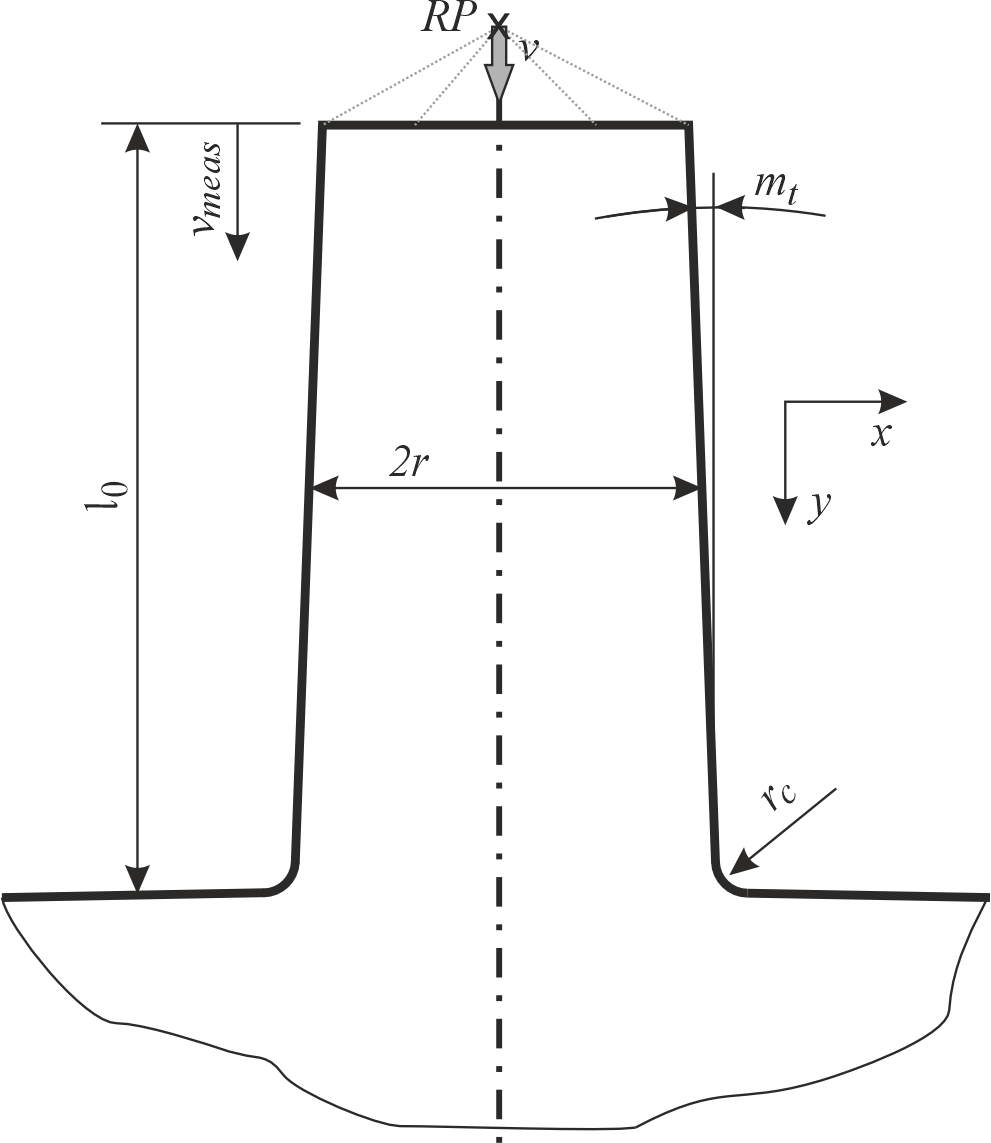}
    \caption{Definition of geometry parameters of micropillars.}
    \label{fig:geo_FEM}
\end{figure}

In previous studies on MPC on wood, local strain fields were not available. As a consequence, only indentor displacements were used for strain calculations, which necessitates a correction of the pillar-base displacement due to the deformation of the bedding, called sink-in. Two widely adopted correction approaches for sink-in during micropillar compression were applied for comparison with strains obtained from DIC: the modified Sneddon correction~\cite{zhang2006design} and Finite Element Method (FEM) simulations. The resulting correction factors are then used to obtain the compressive pillar strain. Without image sequences, only the pillar's initial cross-sectional areas at mid-height are available, and only engineering stresses can be calculated from the measured indentation force (see Fig.~\ref{fig:geo_FEM}). Due to the edge detection, however, one can calculate the real cross-sectional area under the assumption of a circular cross section and consequently true stresses.

The original Sneddon approach~\cite{sneddon1965relation} provides an analytical solution for the penetration depth of an isotropic cylindrical punch into an elastic half-space of the same material due to an external load. The geometrical parameters, such as height ($l_0$), initial mid-height radius ($r$), and fillet radius ($r_c$), are used to define the corrected pillar displacement ($v_{corr}$) from the measured indentor displacement ($v_{meas}$) (see Fig.~\ref{fig:geo_FEM}):
\begin{equation}
    v_{corr} = \xi_S v_{meas} = \frac{-2 l_0 a_c}{A(\nu^2-1)}v_{meas}.
    \label{eq:Sneddon}
\end{equation}
Here $A$ denotes the cross section area, $\nu$ the Poisson's ratio, and $a_c = \eta (r + r_c)$, where $\eta=1.42$ is a modification factor introduced by Zhang et al.~\cite{zhang2006design} from comparison with FEM solutions.

Since the modified Sneddon approach is based on the assumptions of material isotropy and indentation into an elastic halfspace, FEM simulations were performed prior to the experiments to determine a $\xi_{ortho}(MFA)$ factor as a MFA-dependent sink-in correction for the orthotropic cell wall material similar to Refs.~\cite{zhang2006design,cornec2022numerical}. The orthotropic cell wall compliance tensors at \SI{0}{\percent}RH were taken from Ref.~\cite{mora2019mechanical} to match vacuum conditions. Two increasingly realistic FEM models were implemented in Abaqus2021 (see Fig.~\ref{fig:appendix_FEM}): 
\begin{itemize}
    \item A micropillar on a cylindrical substrate with 12 times the pillar radius and three times its length and encastred side edges and bottom, but with orthotropic material corresponding to the S2 layer with different material orientations with respect to the model axis, respectively.
    \item An assembly of tracheids, where the test tracheid with the S2 layer pillar is modelled with tied cell wall layers with respective orthotropic elasticity tensors and material orientation, while surrounding tracheids apply an equivalent single-layer elasticity tensor with discrete material orientation, following the lumen surface. Note that basically two cellular models are required, one for MFA=\SI{0}{\degree} and \SI{20}{\degree} and one for MFA=\SI{70}{\degree} and \SI{90}{\degree} (see Fig.~\ref{fig:sampleprep}).
\end{itemize}
For all models, quadratic fully integrated brick elements are used. A vertical displacement of $v=$\SI{0.5}{\micro\meter} ($\varepsilon_z \approx$\SI{-10}{\percent}) was imposed on a reference point ($RP$ in Fig.~\ref{fig:geo_FEM}) that was kinematically coupled with all degrees of freedom with the top surface of the pillar. This corresponds to the assumption of frictional contact without sliding at the top of the pillar, which is in line with experimental observations~\cite{Clyne_Campbell_2021}. From the ratio between bottom and top displacements of the pillars, the respective correction factor $\xi_{ortho}(MFA)=v_{meas}/v_{corr}$ is determined and summarized in Tab.~\ref{tab:correction_factors}.
\begin{table}[htbp]
\centering
\footnotesize
\caption{Correction factors for different model variants and microfibril angles (MFA).}
\label{tab:correction_factors}
\renewcommand{\arraystretch}{1.8}
\begin{tabular}{lccccc}
\toprule
\textbf{Correction type} & \textbf{Isotropic} & \textbf{00} & \textbf{20} & \textbf{70} & \textbf{90} \\
\midrule
$\xi_S$           & 0.78 & --   & --   & --   & --   \\
$\xi_{ortho}^{Subs.}$    & --   & 0.494 & 0.595 & 0.734 & 0.716 \\
$\xi_{ortho}^{Trach.}$     & --   & 0.450 & 0.558 & 0.687 & 0.662 \\
\bottomrule
\end{tabular}
\end{table}

\begin{figure}[ht]
    \centering
    \includegraphics[width=1\columnwidth]{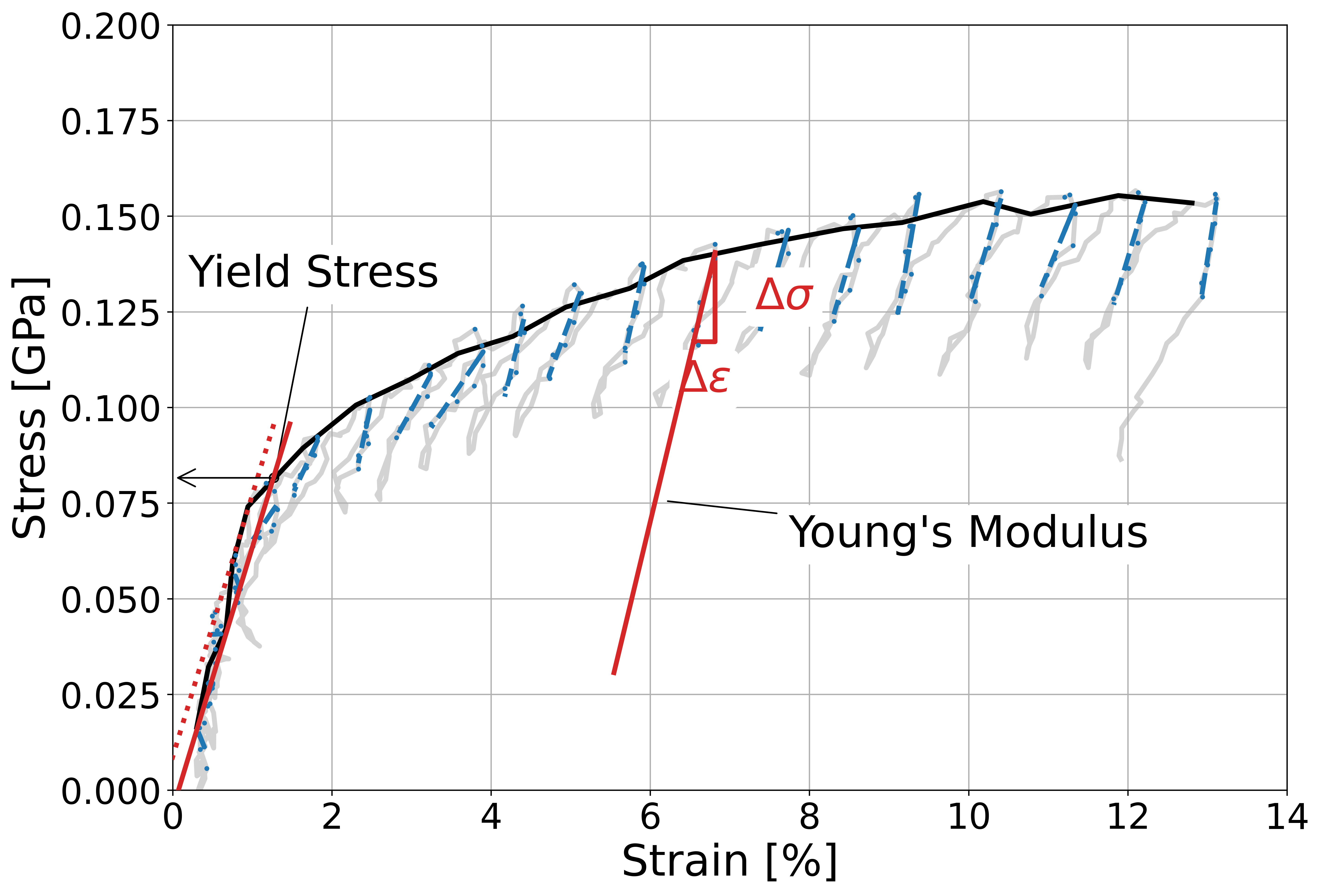}
    \caption{Example of the calculation of mechanical parameters for a pillar with MFA=\SI{70}{\degree}, namely Young's modulus (E), computed for all unloading segments (blue dashed line), and yield stress ($\sigma_\text{yield}$). DIC computed strains are used, with compressive strains being defined as positive.}
    \label{fig:DIC_prot}
\end{figure}

The processed stress and strain data were then used to determine the Young’s modulus ($E$) and yield stress ($\sigma_\mathrm{yield}$) for the different MFAs, as shown in Fig.~\ref{fig:DIC_prot}. Note that compressive strains are defined as positive. The modulus $E$ was calculated as the slope of a linear fit applied to each unloading segment of the stress-strain curve. For the yield stress $\sigma_\mathrm{yield}$, an envelope with the maximum stress per strain value was computed to exclude the unloading segments. A moving regression with a constant window size ($\approx$ \SI{0.5}{\percent}) was applied, and the segment with maximum slope was used to define the yield stress using the \SI{0.2}{\percent} strain offset criterion as done in previous works~\cite{schwiedrzik2016identification}.

\section{Results and Discussions}\label{sec3}   
The success of experiments on the cell wall scale of wood strongly depends on the interplay between specimen preparation quality, the accurate application and measurement of micro-scale forces and displacements, and the robust evaluation of resulting mechanical properties. Together, these factors determine the reliability of insights that are separated into phenomenological observations (Sec.~\ref{sec:phenobs}) and quantitative ones (Sec.~\ref{sec:quantmeasures}) in this manuscript.
\subsection{Phenomenological observations in micropillar testing}\label{sec:phenobs}
\begin{figure}[ht]
    \centering
    \includegraphics[width=1\columnwidth]{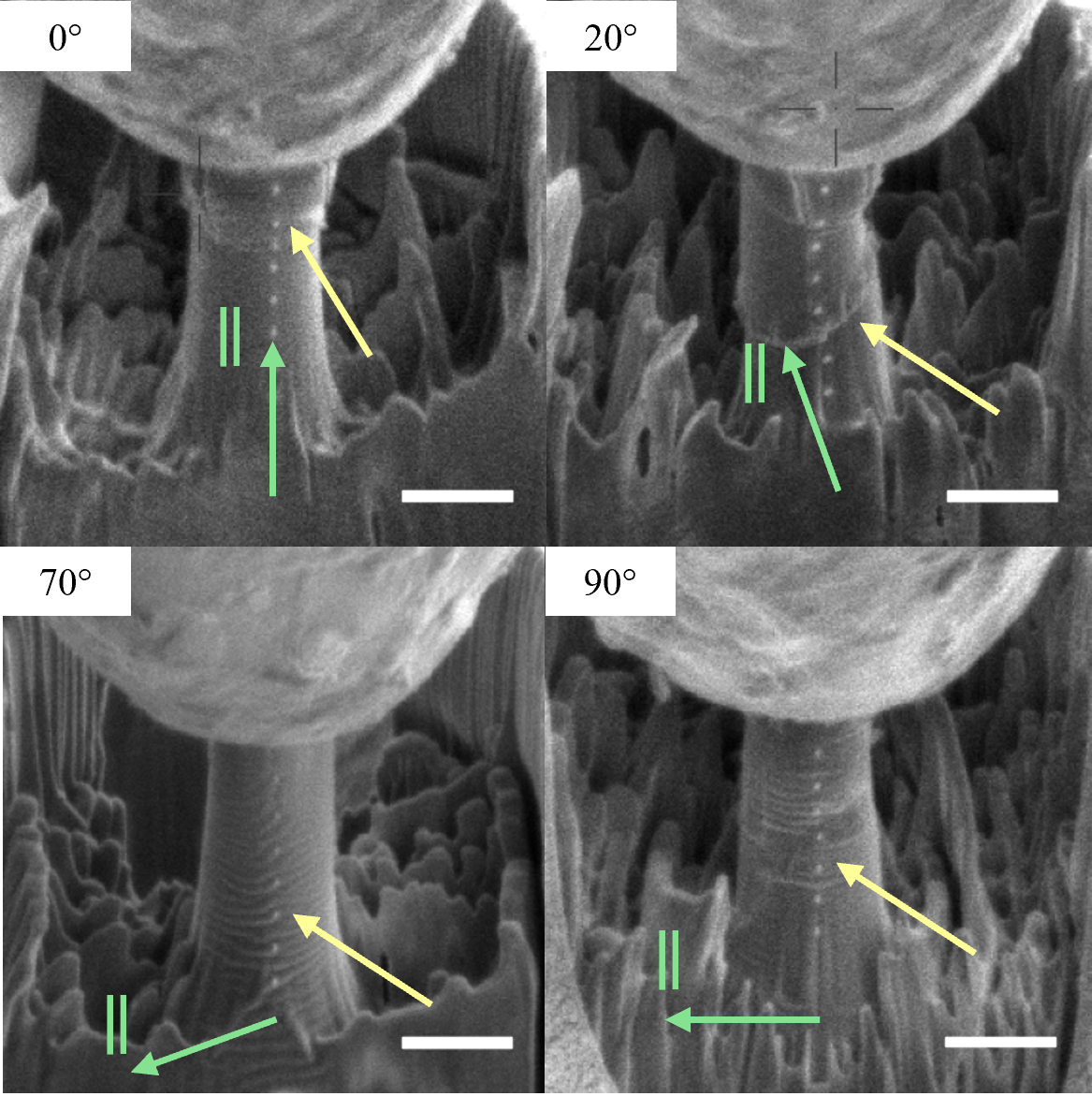}
    \caption{SEM images of observed failure patterns for different orientations highlighted by yellow arrows and the microfibril orientation in green. The scalebar corresponds to \SI{2}{\micro\meter}.}
    \label{fig:pillars_pos}
\end{figure}
An essential factor for increasing reproducibility in micropillar compression tests is the quality of the milled pillars. It is strongly influenced by the surface quality achieved by polishing from which one starts milling with the ion beam, as well as its parameters. Ultramicrotome polishing typically induces shear deformations in the cell wall in the cutting direction~\cite{raghavan2012deformation}, which can result in non-uniform FIB milling. In our experiments, using a microtome step size of \SI{30}{\micro\meter} and wetting the sample surface, similar to that used in histological preparation for wood anatomical sections, smooth surfaces could be realized, leading to a consistent milling process. FIB milling with parameters reported in the range of \SIrange{75}{300}{\pico\ampere} resulted in a grassy irregular surface due to ion channeling effects~\cite{phaneuf2004fib}. The surfaces proved to be much more uniform and smooth when milling with a lower current of \SI{50}{\pico\ampere}, which also resulted in a faster process. Finally, the suitability of the proposed protocol for orienting, polishing, and milling is evident in the morphology of the pillars before (Fig.~\ref{fig:pillars_pre}) and the regularity of failure patterns after testing (Fig.~\ref{fig:pillars_pos}) for all orientations. Nevertheless, one must keep in mind that the material at the base of the milling crater is severely altered and most likely high residual stresses are present due to shrinkage, even resulting in circumferential cracking on the outer rim of the crater as visible in Fig~\ref{fig:pillars_pos}-\SI{0}{\degree}. Sink-in corrections, based on stress-free and intact supporting material, are therefore an idealization.

\begin{figure}[ht]
    \centering
    \includegraphics[width=1\columnwidth]{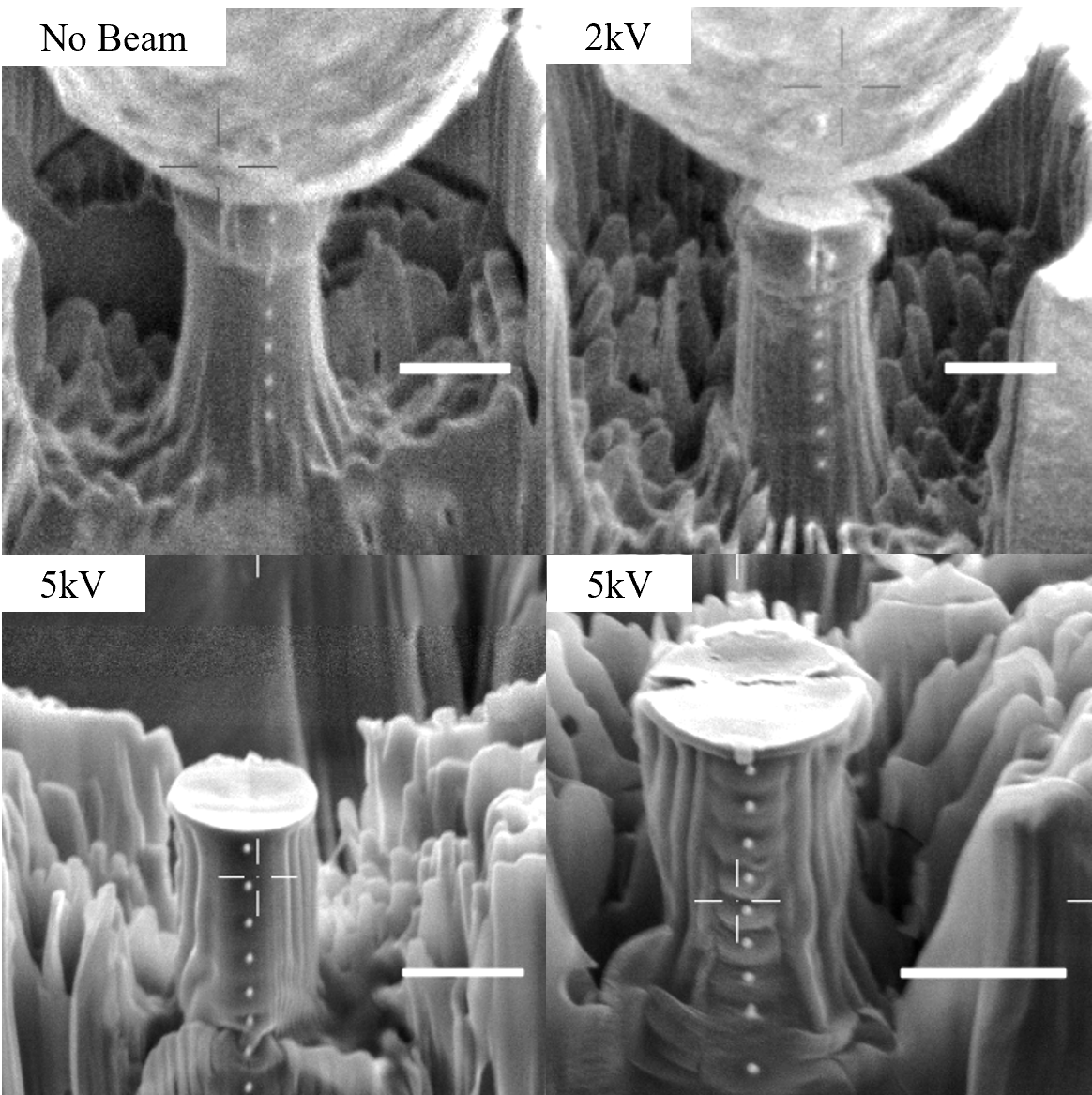}
    \caption{SEM images of observed beam damage on pillars imaged with different acceleration voltages. }
    \label{fig:comparison_pic}
\end{figure}

The material behavior can also be altered during testing due to electron beam damage caused by the SEM. One observes beam-induced shrinkage of the pillar, accompanied by the formation of a vertically shriveled surface morphology. The high sensitivity of the wood cell wall to electron irradiation results from localized energy deposition, leading to localized heating and the disruption of chemical bonds~\cite{sezen2011investigation,grubb1974radiation}. Figure~\ref{fig:comparison_pic} compares representative pillars subjected to different imaging protocols described in Sec.~\ref {sec:compprot}. The unscanned pillar retained a smooth surface and well-defined geometry, whereas the pillar imaged at \SI{2}{\kilo\volt} displayed a rough, skin-like texture. However, its deformation and failure features remained identifiable. The pillars scanned at \SI{5}{\kilo\volt} exhibited extensive electron beam damage, characterized by pronounced shrinkage and the absence of visible failure marks — a condition observed consistently across all MFA orientations at this scanning protocol. It is important to note that earlier micropillar tests on wood may have unknowingly introduced varying degrees of beam-induced degradation~\cite{adusumalli2010deformation, schwiedrzik2016identification, klimek2020micromechanical, zhang2010characterizing}.

The composite nature of the wood cell wall introduces additional complexity in its thermal transport behavior. This was particularly evident in high-MFA pillars ($70^{\circ}$ and $90^{\circ}$), which showed disproportionately higher degradation compared to low-MFA counterparts. We attribute this to the anisotropic thermal conductivity of cellulose microfibrils~\cite{wang2022enhanced,pan2021high}, where transverse fibril orientations limit heat transport into the bulk or the indentor, promoting internal heat accumulation and thermal degradation within the pillar. These observations confirm that beam-induced thermal and chemical effects can collectively compromise the structural and mechanical integrity of wood micropillars and, therefore, require special attention, as they cannot be entirely avoided. 

The emerging surface patterns during compressive loading allow conclusions on the failure mechanisms, which strongly depend on the MFA (see Fig.~\ref{fig:pillars_pos}). The two pillars with MFAs of \SI{0}{\degree} and \SI{20}{\degree} exhibit kink bands formed perpendicular to their respective microfibril orientations. In contrast, the two pillars with MFAs of \SI{70}{\degree} and \SI{90}{\degree} exhibit failure patterns characterized by the compressive collapse of fibrillar layers. It appears that FIB milling has resulted in a thin "skin" with altered behavior on the pillars, which prevents the localization in a single dominant shear plane, as previously observed in metals~\cite{el2009effects}. Furthermore, pictures indicate a correct alignment between the compression direction and the ultrastructural orientation of the S2 cell wall.

\subsection{Mechanical behavior of S2 micropillars}\label{sec:quantmeasures}
\begin{figure}[ht]
    \centering
    \includegraphics[width=1\columnwidth]{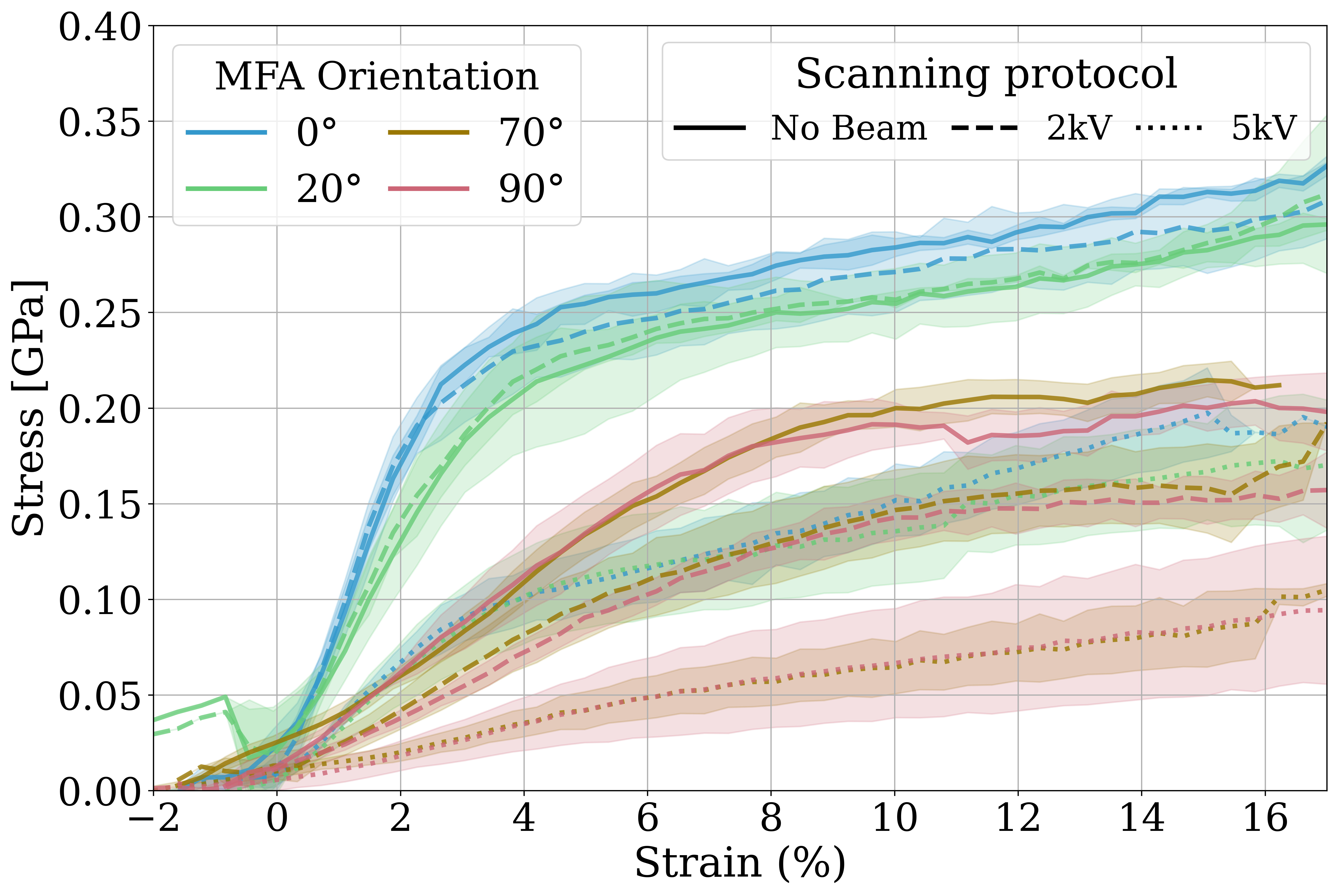}
    \caption{Average stress envelope and confidence interval of stress-strain curves for micropillars at \SI{0}{\degree}, \SI{20}{\degree}, \SI{70}{\degree}, and \SI{90}{\degree} under three scanning protocols. Each curve averages at least 4 micropillars.}
    \label{fig:comparison}
\end{figure}

Beam-induced damage during SEM imaging can also significantly impact the mechanical behavior of the cell walls and, consequently, the results of the MPC tests. Figure~\ref{fig:comparison} presents the mechanical behavior of pillars with four different MFAs measured with three distinct imaging protocols. For all orientations, pillars tested without SEM scanning exhibited higher stiffness and sustained greater loads, indicating reduced material degradation. The introduction of scanning, particularly at higher acceleration voltages (\SI{5}{\kilo\volt}), led to a consistent reduction in mechanical performance.  Although amplified by beam-induced damage, particularly under \SI{5}{\kilo\volt} imaging conditions, negative lateral strains were consistently observed across all MFA orientations (see Video Fig.~\ref{fig:video}) (and video). While this behavior could suggest a negative Poisson’s ratio, experimental artifacts make it difficult to determine whether these measurements reflect an intrinsic material response or result from local structural degradation. Notably, a relaxation of the negative lateral strain was observed once the pillar was unloaded entirely, which could indicate either a genuine auxetic behavior of the material or the release of compressive transverse stresses introduced by beam damage-induced shrinkage. However, lower acceleration voltages (\SI{2}{\kilo\volt}) effectively mitigated these effects, offering a potential compromise between image quality and structural preservation.

\begin{figure}[ht]
\begin{center}
\begin{minipage}[b]{1\columnwidth}
    \centering
    \includegraphics[width=1\columnwidth]{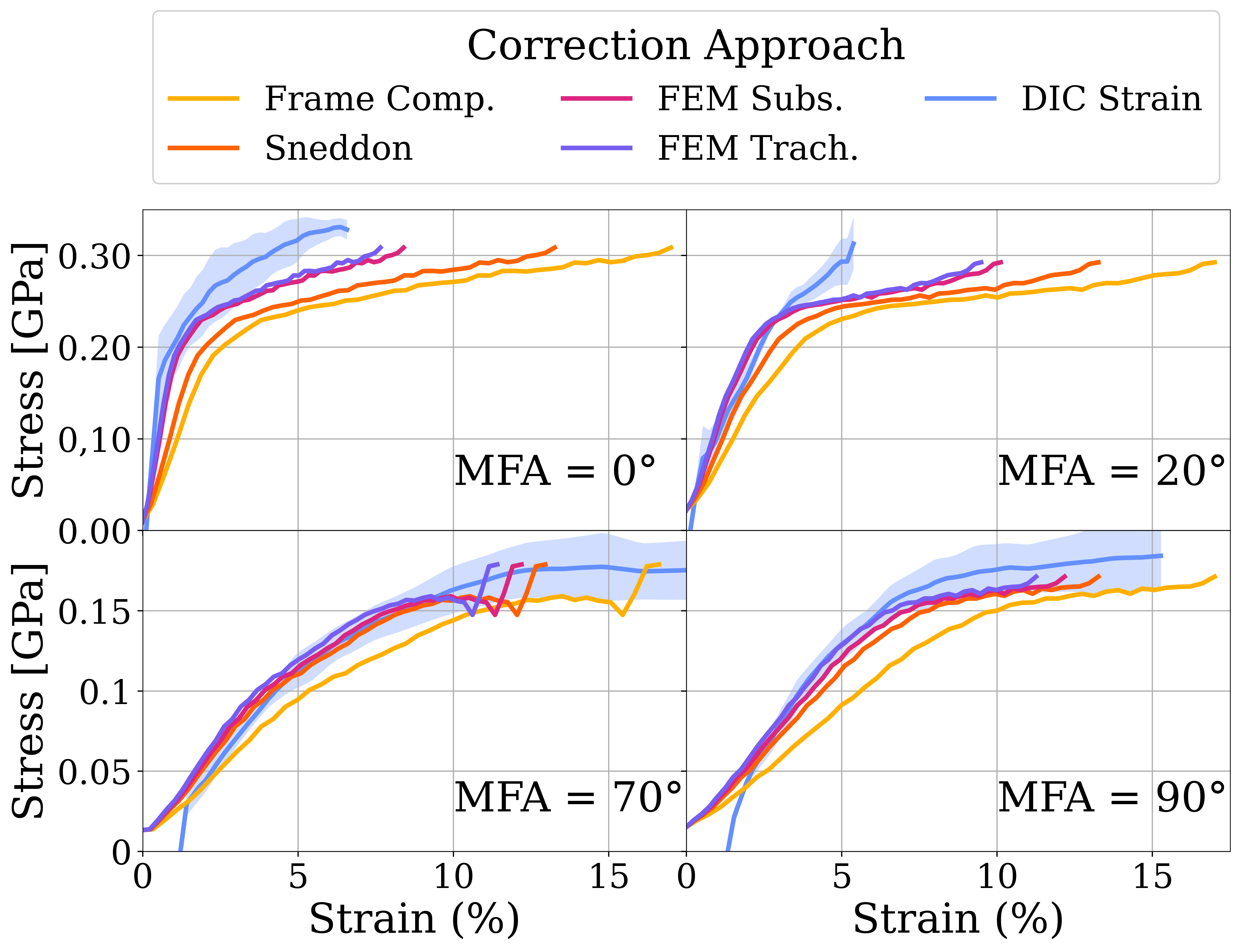}
    \end{minipage}
    \caption{Comparison of different sink-in correction approaches. Note that identical initial strain offset corrections are applied as described for Fig.~\ref{fig:comparison}. The confidence intervals for non-DIC curves are omitted for clarity but visible in Fig.~\ref{fig:comparison}.}
    \label{fig:DIC_res}
\end{center}
\end{figure}

Digital image correlation (DIC) consistently measured lower strain values than the indentor displacement-based strains. The performance of the different correction methods for sink-in is shown in Fig.~\ref{fig:DIC_res}. The low-MFA pillars—curves for MFA \SI{0}{\degree} and \SI{20}{\degree}, respectively—exhibited the most significant discrepancies between DIC-based and indentor-based strain, highlighting the influence of sink-in effects. In contrast, the high-MFA pillars—curves for MFA \SI{70}{\degree} and \SI{90}{\degree}—showed a smaller difference between the two methods, with DIC strains representing about \SI{35}{\percent} of the total deformation. Note that, due to the adaptive ROI, zones with localized damage are excluded from the DIC strain calculation. As a matter of fact, the sink-in correction with orthotropic material on the tracheid using the indentor displacement gets reasonably close to the more accurate DIC strain measurement.

The average mechanical parameters derived from the tests are summarized in Table~\ref{tab:emod_yield_column} and compared with previously reported values for the S2 cell wall layer. When imaging the pillars at an acceleration voltage of \SI{2}{\kilo\volt}, the Young’s modulus obtained via DIC for MFA = \SI{0}{\degree} was E = \SI{42(3)}{\giga\pascal}, while recent model predictions are of E = \numlist{55;69}\SI{}{\giga\pascal}~\cite{salmen2004micromechanical,mora2019mechanical}. For MFA = \SI{90}{\degree}, E = \SI{7(1)}{\giga\pascal}, while recent model predicts E = \numlist{8}\SI{}{\giga\pascal}~\cite{mora2019mechanical}. These results represent the closest match between micromechanical testing and modeling reported to date. DIC-based strain measurements at lower acceleration voltages yielded stiffness values up to \SIrange{35}{120}{\percent} higher than those calculated directly from indentor displacements, confirming that global displacement measurements tend to underestimate the true material response. However, when corrected with a realistic sink-in model, the global displacement values approximate those from DIC. The high degradation of pillars scanned at an acceleration voltage of \SI{5}{\kilo\volt} resulted in DIC strain values similar to the non-corrected indentor displacements, and a Young's modulus comparable to values reported in literature~\cite{klimek2020micromechanical}, of E = \SI{8(3)}{\giga\pascal} for MFA = \SI{0}{\degree}. Across all MFA orientations, modulus values followed the expected decreasing trend with increasing MFA. 

\begin{table}[htbp]
\centering
\caption{Mechanical properties (\(E\) and \(\sigma_{\mathrm{yield}}\)) for different MFA orientations and imaging protocols. All values are presented in \SI{}{\giga\pascal}. The correction factors with realistic tracheid geometry, $\xi_{ortho}^{Trach.}$, are applied to values with subscript FEM. In the literature values, MPC stands for Micropillar Compression and NI for Nanoidentation.}
\renewcommand{\arraystretch}{1.2}
\footnotesize
\begin{tabular}{llcccc}
\toprule
\textbf{MFA (\SI{}{\degree})} & \textbf{Protocol} & \textbf{\(E_{\mathrm{DIC}}\)} & \textbf{\(E_{\mathrm{FEM}}\)} & \textbf{\(\sigma_{\mathrm{yield, DIC}}\)}  & \textbf{\(\sigma_{\mathrm{yield, FEM}}\)} \\
\midrule
0  & No Beam & -- & \textbf{38(4)} & -- & \textbf{0.23(1)} \\
0  & 2kV     & \textbf{42(3)} & 37(3) & 0.18(4) & 0.21(2) \\
0  & 5kV     & 8(3)  & 23(4) & -- & 0.07(1) \\
20 & No Beam & -- & 26(3) & -- & 0.20(4) \\
20 & 2kV     & 26(5) & 25(4) & 0.14(3) & 0.20(3) \\
20 & 5kV     & 4(1)  & 15(3) & -- & 0.07(1) \\
70 & No Beam & -- & 9(2)  & -- & 0.15(1) \\
70 & 2kV     & 11(1) & 7(2)  & 0.10(2) & 0.09(2) \\
70 & 5kV     & 1(1)  & 4(1)  & -- & 0.05(1) \\
90 & No Beam & -- & \textbf{7(1)}  & -- & 0.14(1) \\
90 & 2kV     & \textbf{7(1)}  & 6(2)  & 0.09(2) & 0.12(2) \\
90 & 5kV     & 1(1)  & 4(2)  & -- & 0.05(2) \\
\midrule
\multicolumn{6}{l}{\textbf{Literature}} \\
\midrule
\textbf{MFA (\SI{}{\degree})} & \textbf{\(E\)} & \textbf{\(\sigma_{\mathrm{yield}}\)} & \textbf{Method} & \textbf{Source} \\
\midrule
0--20 & 8      & 0.18 & MPC      & \cite{klimek2020micromechanical,schwiedrzik2016identification} \\
0--20 & 26         & --   & NI   & \cite{jager2011identification} \\
0     & \textbf{69, 55} & \textbf{0.24} & Modelling        & \cite{mora2019mechanical, salmen2004micromechanical, schwiedrzik2016identification} \\
20    & 35         & --   & Modelling         & \cite{salmen2004micromechanical} \\
90    & \textbf{8}          & --   & Modelling         & \cite{mora2019mechanical} \\
40    & --         & 0.05 & MPC       & \cite{schwiedrzik2016identification} \\
40    & --         & 0.04 & Modelling         & \cite{schwiedrzik2016identification} \\
\bottomrule
\end{tabular}
\label{tab:emod_yield_column}
\end{table}

The yield stress values $\sigma_\mathrm{yield}$ obtained in this study were in good agreement with both experimental~\cite{schwiedrzik2016identification,klimek2020micromechanical} and modeled~\cite{schwiedrzik2016identification} data. In contrast to modulus, yield stress is more reliably extracted from the load-displacement curve and is less affected by noise in the early loading stages. However, beam-induced damage—particularly under high-voltage SEM imaging—reduces the overall load-bearing capacity of the pillars, leading to underestimation of $\sigma_\mathrm{yield}$ (see Tab.~\ref{tab:emod_yield_column}).
\section{Conclusion}
\begin{figure*}[ht]
    \centering
    \includegraphics[width=1\linewidth]{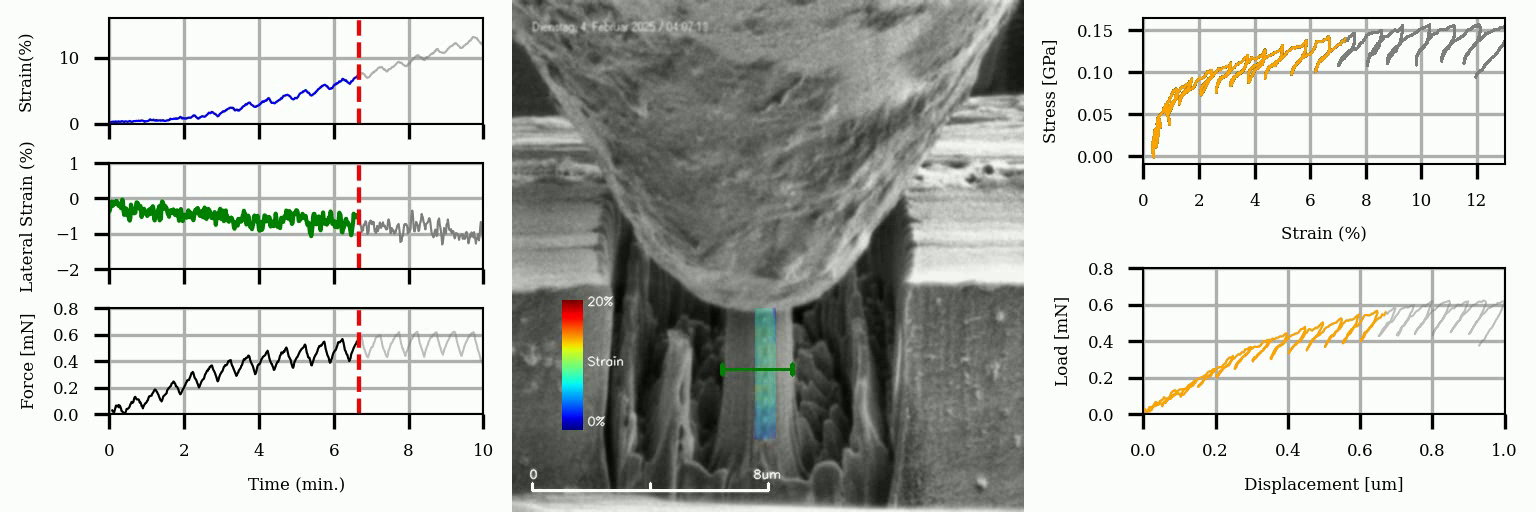}
    \caption{Frame of video of DIC measurement during compression test. The longitudinal and lateral strains and raw load measurements are on the left. In the center, the compression video is overlaid with the DIC strain field and the lateral strain measurement. On the right, the stress-DIC strain and the load-displacement curves are shown.}
    \label{fig:video}
\end{figure*}

This study elaborates that a critical discussion on the reliability and reproducibility of micropillar (MP) compression as a micromechanical technique is of fundamental importance for a mechanical characterization of soft, anisotropic, and organic materials of the ultrascale, e.g., the S2 layer in wood. To do so, MPs aligned in different orientations relative to the wood cell wall material, e.g., different Microfibrillar Angles (MFA), were tested in vacuum conditions in a Scanning Electron Microscope (SEM) chamber. New insights could be gained by introducing a direct, optical strain measurement approach with digital image correlation (DIC) based on a row of microdots that were deposited along the micropillar surfaces. A typical video of the experiment with evaluated strains and stresses is given in Fig.~\ref{fig:video}. By comparing local strains with indentor displacements, statements on the suitability of correction approaches of pillar penetration into the substrate could be made. Commonly, models for isotropic materials are used, even though materials are orthotropic and substrates are not infinite in the case of the wood cell wall. With simulations of FEM models of increasing realism, like orthotropic halfspace and further orthotropic cellular assembly, the suitability of such sink-in corrections on indentor displacements could be shown.

When comparing the sink-in corrected strain values to the DIC measurements, it was observed that standard isotropic models underestimate the penetration depth of MPs in anisotropic materials. Moreover, the commonly used half space as substrate in FEM simulations doesn't capture the border effect of microstructured materials, such as the lumen effect on the cell wall deformation. The strain correction approximated the DIC-measured values when applying realistic geometries and orthotropic material properties. However, the sink-in-based corrections are linear and cannot account for nonlinear behaviors such as plasticity and failure. As a result, the DIC-measured mechanical parameters were higher and closer to the state-of-the-art micromechanical models than any sink-in correction. As a consequence, results from bio-indentors without the possibility to monitor local strains are possible, but require an advanced sink-in correction.

\begin{figure}[ht]
    \centering
    \includegraphics[width=1\columnwidth]{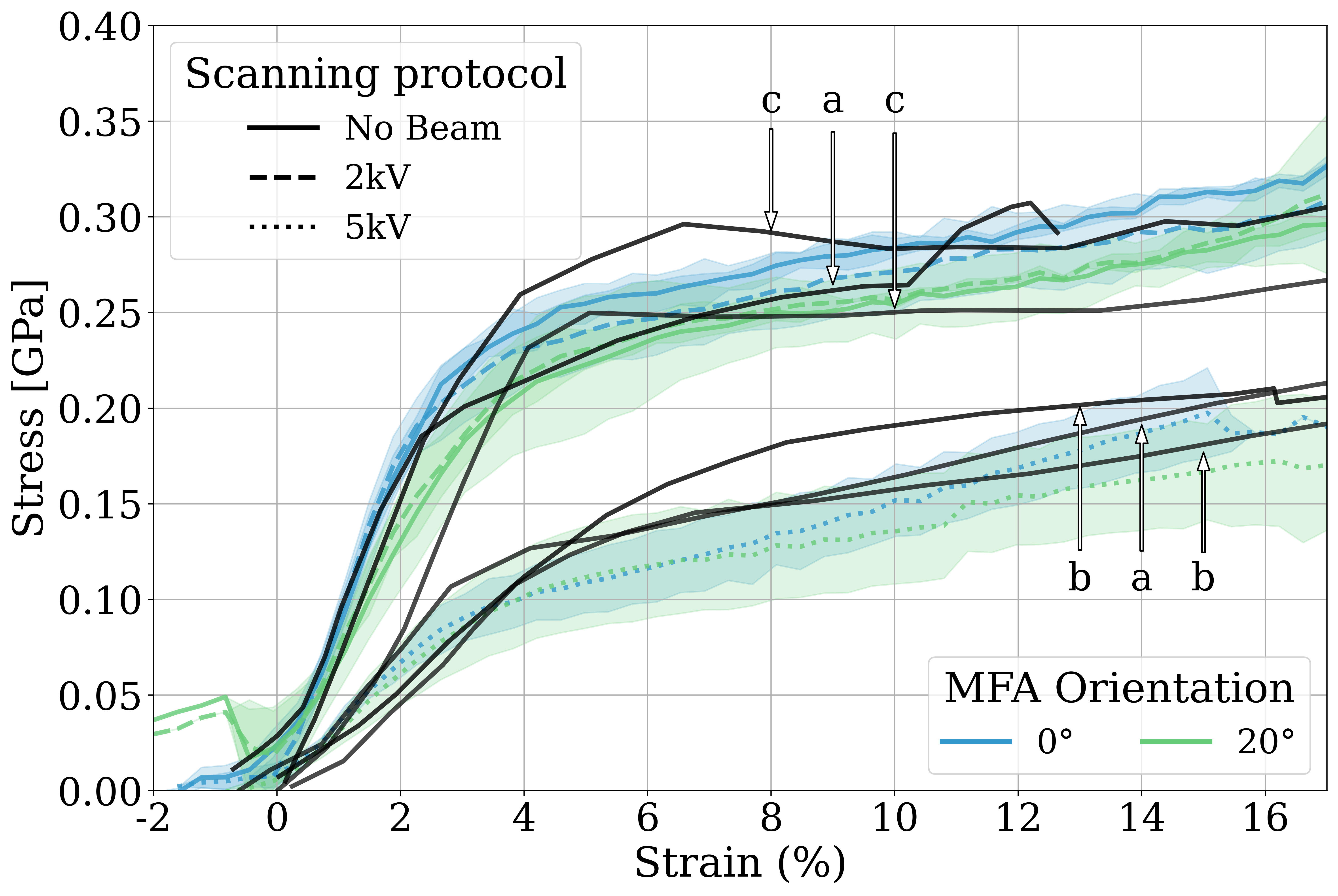}
    \caption{Stress-strain curves for different scanning protocols and comparison with literature (black solid lines: (a)\cite{schwiedrzik2016identification}; (b)\cite{adusumalli2010deformation}; (c)\cite{klimek2020micromechanical}). All curves presented are from micropillars with microfibrillar angle between \numlist{0;20}\SI{}{\degree}.}
    \label{fig:comparison_lit}
\end{figure}

Despite enabling access to local strain fields, digital image correlation (DIC) requires continuous scanning during testing, which leads to significant electron beam-induced damage. Different electron beam acceleration voltages were used during imaging to identify possible damage caused by electron-wood interactions. 
This degradation was evident when comparing the mechanical response of pillars with identical orientations subjected to different imaging protocols. Across all tested MFA orientations, a strength reduction of at least \SI{50}{\percent} was observed in pillars continuously scanned at high acceleration voltages (\SI{5}{\kilo\volt} for \SI{15}{\minute}) compared to unscanned ones. When the present results are compared with previously published data, it becomes apparent that the wide scatter reported in earlier studies spans the full range of responses, from minimally degraded to severely degraded pillars, see Fig.~\ref{fig:comparison_lit} and references therein. The sample preparation and testing protocol presented in this study helped clarify the sources of data scatter reported in previous works, thereby improving the agreement between experimental micromechanical testing and modeling at the wood cell wall scale. These findings indicate that uncontrolled electron beam exposure in previous work likely caused varying degrees of material degradation, representing the primary source of the observed variability in mechanical response.

Despite the challenges, the proposed methodology effectively captured the mechanical responses of the wood cell wall for different MFA orientations. The measured values, E=\SI{42(3)}{\giga\pascal} and $\sigma_\mathrm{yield} =$ \SI{0.19}{\giga \pascal} for MFA=\SI{0}{\degree}, are the closest experimental estimates to recent micromechanical model predictions (E= \numlist{55;69}\SI{}{\giga\pascal}~\cite{hassani2015rheological,salmen2004micromechanical}, $\sigma_\mathrm{yield} =$ \SI{0.18}{\giga \pascal}~\cite{schwiedrzik2016identification}). The value for MFA = \SI{90}{\degree} with E = \SI{7(1)}{\giga\pascal} even matched the model predictions (E = \numlist{8}\SI{}{\giga\pascal}) ~\cite{mora2019mechanical}. Additionally, failure patterns were consistent with fibril orientation, with fibril-aligned kink bands observed at low MFAs and compressive collapse of fibrillar layers at high MFAs. While the current setup limits the extraction of viscoelastic or moisture-dependent properties, continued advances in FIB precision and SEM imaging may enable such analyses in future work.

\section{Acknowledgements}
We acknowledge the financial support from the Swiss National Science Foundation under SNF grant 200021 192186,” Creep behavior of wood on multiple scales.”. We want to thank Dr. Daniele Casari, Dr. Tijmen Vermeij, and Dr. Fedor Klimashin from Empa-Thun for training and helping with the experimental apparatus used in this work. We want to thank Prof. Johann Michler for providing us with access to the EMPA Thun facility and equipment, which made this work possible. We want to thank Dr. Maximiliam Ritter for the SAXS measurements and the discussions about electron beam damage, together with Dr. Jonas Garemark and Ronny Kürsteiner.
\section{Declaration of generative AI and AI-assisted technologies in the writing process}
During the preparation of this work, the author(s) used ChatGPT 4o to improve the clarity, flow, and readability of the text. After using this tool/service, the authors reviewed and edited the content as needed and take full responsibility for the publication's content.
\section{Appendix A - Finite Element Methods structure}
\begin{figure*}[ht]
    \centering
    \includegraphics[width=0.8\linewidth]{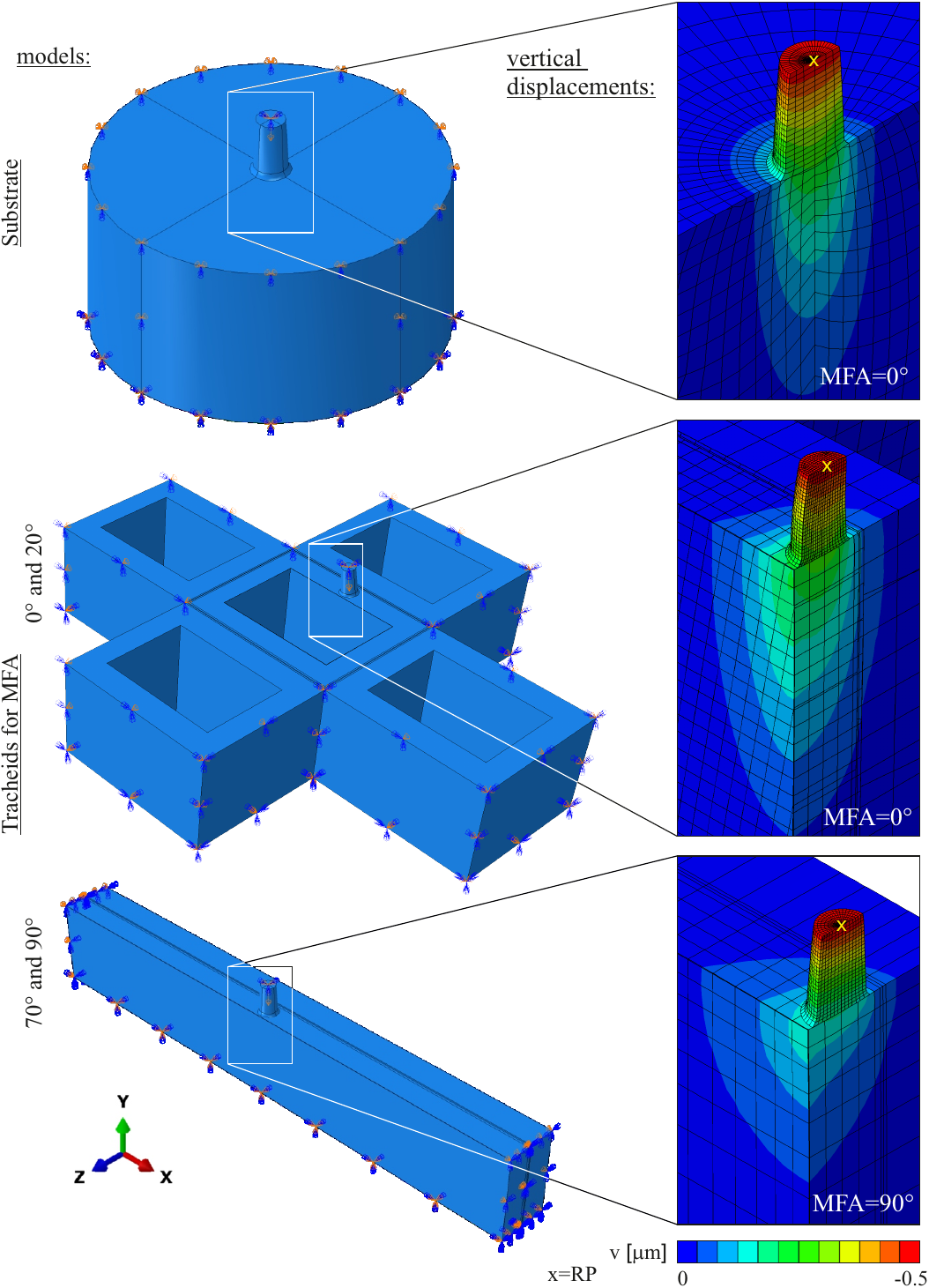}
    \caption{Geometries and boundary conditions used for FEM simulations, the reference point where vertical displacements were imposed is shown as "RP". The same color scale is used for all figures.}
    \label{fig:appendix_FEM}
\end{figure*}

The simulated geometries in Finite Element Methods are shown in Fig.~\ref{fig:appendix_FEM}. In the first row is the simpler geometry where the substrate is adopted as a cylinder, but the orthotropic material properties are applied in different orientations representing the different microfibrillar angles (MFA): \numlist{0;20;70;90}\SI{}{\degree}. The second row shows the geometries as an assembly of tracheids, used for pillars with MFA =\numlist{0;20}\SI{}{\degree}. The tracheid where the pillar is placed has a layered structure, consisting of a middle lamella, primary wall, and secondary layers S1, S2, and S3. The neighboring tracheids have an equivalent material property representing all layers. The third row shows the geometries as an assembly of tracheids, used for pillars with MFA =\numlist{70;90}\SI{}{\degree}.
\label{app1}

\clearpage
\newpage
\bibliographystyle{elsarticle-num}
\bibliography{lit}
\end{document}